\documentclass{PoS}
\usepackage{amsmath}
\usepackage{braket}
\usepackage{psfrag}
\usepackage[usenames]{color}
\makeatletter
\let\default@color\current@color
\makeatother

\DeclareRobustCommand{\JCCvec}[1]{\boldsymbol {\mathbf{#1}}}
\newcommand\3[1]{\JCCvec{#1}}

\DeclareRobustCommand*\diff[2][]{%
   \mathop{
         \mathrm{d}^{#1}
      \mskip-0.2\thinmuskip
   #2}\nolimits
}

\newcommand{\trace}{\mathop{\mathrm{Tr}}}

\newcommand \T[1]{\3{#1}_T}

\providecommand{\VCeps}[3]{%
    \raisebox{-0.5\height}{#2[#1]{#3}}%
}

\definecolor{brown}{cmyk}{0,0.81,1,0.60}
\definecolor{darkgreen}{rgb}{0,0.6,0}
\definecolor{greyout}{rgb}{0.5,0.5,0.5}
\definecolor{orange}{rgb}{1.0,0.7,0}

\newcommand \black  {\color{black}}

\newcommand \blue  {\color{blue}}
\newcommand \brown {\color{brown}}
\newcommand \darkgreen {\color{darkgreen}}

\newcommand \red   {\color{red}}

\newcommand \hard   {\darkgreen }
\newcommand \proton {\blue}

\newcommand \jet    {\brown}

\newcommand\standardsubs{%
   \psfrag{e}{\small \hard $e$}
   \psfrag{q}{\small \hard $q$}
   \psfrag{k}{\small \red $k$}
   \psfrag{k'}{\small \jet $k'$}
   \psfrag{k+q}{\small \jet $k+q$}
   \psfrag{p'}{\small \jet $p'$}
   \psfrag{p}{\small \proton $p$}
}

\title{Rapidity divergences and valid definitions of parton densities }

\ShortTitle{Rapidity divergences and valid definitions of parton densities }

\author{\speaker{John COLLINS}\\
        Physics Department, Pennsylvania State University, USA\\
        E-mail: \email{collins@phys.psu.edu}
}

\abstract{%
  Rapidity divergences occur when parton densities in a gauge theory
  are defined in the most natural way, as expectation values of
  partonic number operators in light-front quantization.  I review
  these and other related divergences, and show how the definitions of
  parton densities can be modified to remove the divergences.  A
  modified definition is not only essential for many phenomenological
  applications of QCD, but also concerns the treatment of parton
  densities in non-perturbative approaches.  The necessity of
  modifications in the definition of a parton density also entails
  corrections in the formulation of light-front quantization for gauge
  theories.  }

\FullConference{LIGHT CONE 2008 Relativistic Nuclear and Particle Physics\\
		 July 7-11, 2008\\
		 Mulhouse, France}

\begin{document}

\section{Introduction}

Parton densities are central to much phenomenology of scattering in
QCD.  Hard-scattering factorization theorems represent cross sections
as convolutions of perturbatively calculable factors and of parton
densities (and related quantities like fragmentation functions).
Predictions are made on the basis of perturbative calculations because
of the universality of parton densities between different reactions.

Natural candidate definitions of parton densities are obtained from
elementary treatments of the parton model, and they can be formulated
as expectation values of number operators in light-front
quantization.  
Unfortunately, when applied in a gauge theory like QCD, these
definitions suffer from divergences where the rapidities of some
gluons goes to infinity.

This talk reviews how rapidity divergences and certain other
divergences arise.  It summarizes their impact on correct definitions
of parton densities, on phenomenology, and on fundamental issues in
light-front quantization.  Naturally, a valid derivation of a
factorization property requires a valid definition of parton densities
that is matched to the factorization property.  Moreover, to allow
non-perturbative methods in QCD to be used to estimate parton
densities, operator definitions of parton densities are needed that
can be taken literally.

\section{Parton densities from parton model}

In view of the many complications in full QCD, we first recall how
parton densities arise in the parton model.  This leads to a natural
definition of a parton density, thereby giving a convenient conceptual
landmark.  The results in QCD can be regarded, not as overthrowing the
parton model but as modifying and distorting it, while preserving much
of an overall intuitive framework.

We consider DIS, $ep\to eX$, and use light-front coordinates:
$k^{\pm}=(k^0\pm k^z)/\sqrt2$.  We work in the Breit frame, where the
momenta of the target and the virtual photon are 
$p^\mu = \left( p^+, \, M^2/(2p^+), \, \T{0} \right)$
and 
$q^\mu = \left( -xp^+, \, Q^2/(2xp^+), \, \T{0} \right)$.
In the parton model, valid in certain model field theories, we assume
that, to leading power in $\text{mass}/Q$, there is dominance by
handbag diagrams
\begin{equation}
\standardsubs
\label{eq:handbag}
  F_2  
~\simeq~ 
      \int \frac{\diff[4]{k}}{(2\pi)^4} \quad
      \VCeps{scale=0.4}\includegraphics{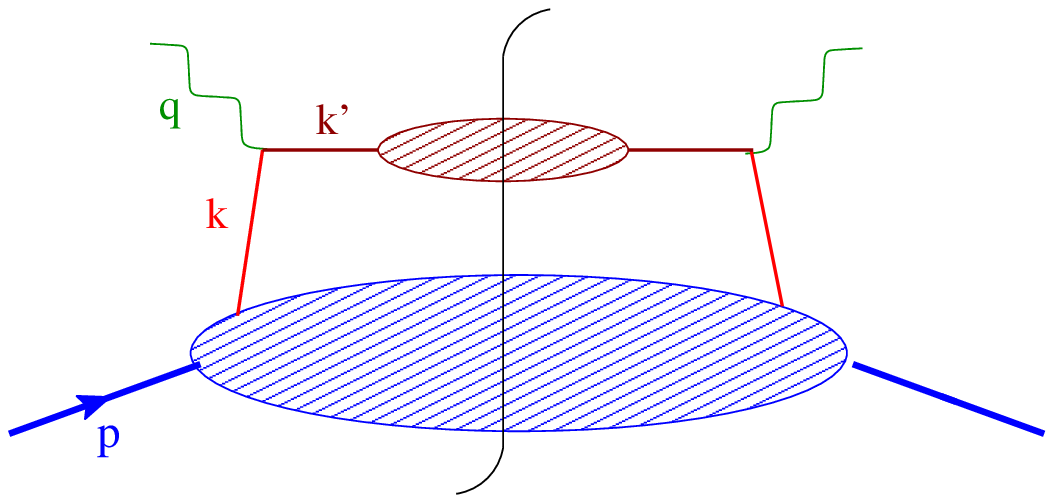}
\end{equation}
with the virtualities and transverse momentum of the quark
lines being limited.

This leads to well-known scaling formulae for structure functions,
such as $F_2=\sum_j e_j^2 xf_j(x)$.  Here $f_j$ is the density of a quark
of flavor $j$ defined graphically by
\begin{equation}
\standardsubs
\label{eq:pdf1.diagram}
   f_j(x)
\stackrel{\text{parton model}}{=}
   \trace \frac{\gamma^+}{2}
     \int \frac{ \diff{k^-} \diff[2]{\T{k}} }{ (2\pi)^4 }
      \quad
      \VCeps{scale=0.32}\includegraphics{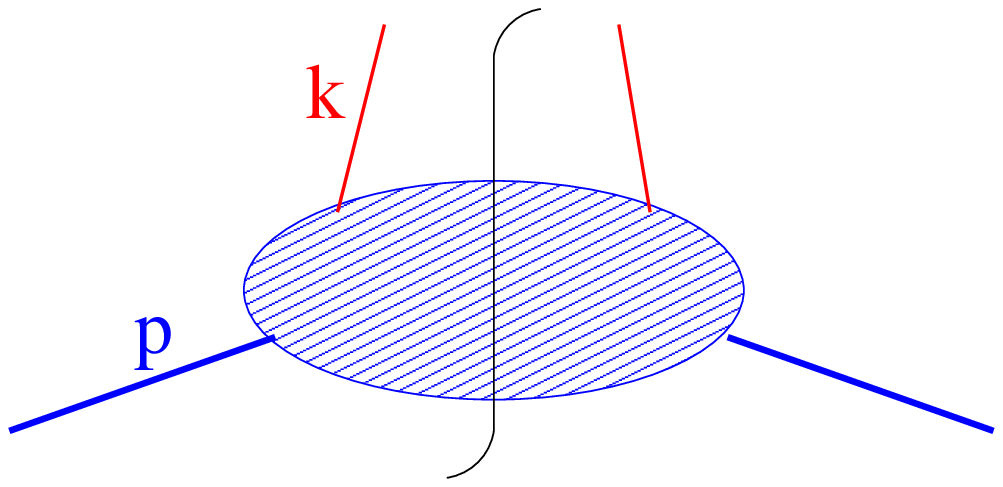}
\end{equation}
with $k^+=xp^+$, which relates the momentum of the internal quark to
the experimentally measurable quantity $x$.  The corresponding
operator formula is
\begin{align}
  \label{eq:pdf1.op}
  f_j(x) 
\stackrel{\text{parton model}}{=}
  \int \frac{ \diff{w^-} }{ 2\pi } \,
    e^{ -ix P^+w^- }
    \braket{P| \,
            \overline{\psi}_j(0,w^-,\T{0}) \,
            \frac{\gamma^+}{2} \,
       \psi_j(0) \,
    |P}_{\text{c}} ,
\end{align}
where $\psi_j(w^+,w^-,\T{w})$ is the quark field, and the subscript ``c''
denotes that only connected graphs are to be included.  The two fields
in Eq.\ (\ref{eq:pdf1.op}) have a light-like separation, because of
the integral over all $k^-$ and $\T{k}$ in Eq.\ (\ref{eq:pdf1.diagram}).

Some other processes, e.g., semi-inclusive DIS, are sensitive to
parton transverse momentum.  For these, we use an unintegrated parton
density defined by removing the integral over $\T{k}$, to give
\begin{align}
  \label{eq:pdf2}
f_j(x,\T{k}) 
\stackrel{\text{parton model}}{=}
  \int \frac{ \diff{w^-} \diff[2]{\T{w}} }{ (2\pi)^3 } \,
    e^{ -ix P^+w^- + i \T{k}\cdot\T{w} }
    \braket{P| \,
            \overline{\psi}_j(0,w^-,\T{w}) \,
            \frac{\gamma^+}{2} \,
       \psi_j(0) \,
    |P}_{\text{c}} .
\end{align}

An interpretation is made by using light-front quantization.  
Expanding the fields in annihilation and creation operators gives
\begin{equation}
  \label{eq:pdf3}
f_j(x,\T{k}) 
\stackrel{\text{parton model}}{=}
\sum_\lambda \frac{ \braket{p| ~ b_{k,\lambda,j}^\dag b_{k,\lambda,j} ~ |p}  - \mbox{vacuum
    expectation value}  }
                   { \braket{p|p} }
              \times \frac{1}{ 2x (2\pi)^3 },
\end{equation}
where $b_{k,\lambda,j}$ is the annihilation operator for a quark of flavor
$j$, helicity $\lambda$, plus-momentum $xp^+$ and transverse momentum
$\T{k}$.  To the extent that the parton model is exactly correct, a
parton density is therefore the number density of partons of the given
momentum and flavor.

\section{Wilson lines in parton densities}

\subsection{Light-cone gauge}

To apply light-front quantization in a gauge theory, it is natural to
use the light-cone gauge $A^+=0$.  For partons collinear to the target,
this gauge leads to a treatment that is the same as in the parton
model.  So it is very natural to try to continue to use Eq.\
(\ref{eq:pdf2}), or equivalently (\ref{eq:pdf3}).
However, such a definition gives problems from kinematic
regions other than the collinear-to-target region.  

\begin{figure}
  \centering
  \begin{tabular}{c@{\hspace{1cm}}c@{\hspace{1cm}}c}
      \includegraphics[scale=0.5]{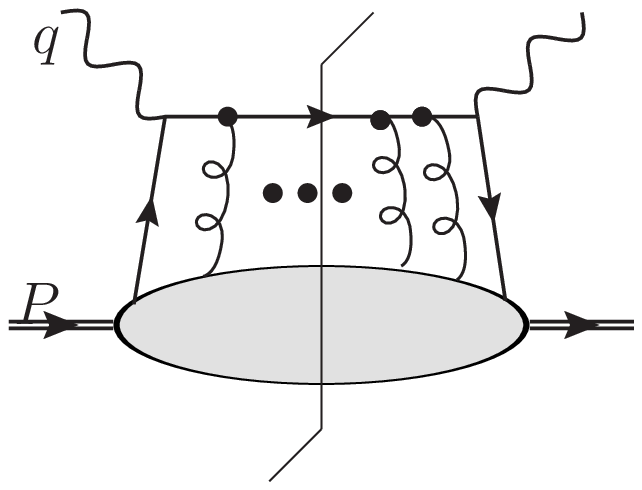}  
    &
      \includegraphics[scale=0.5]{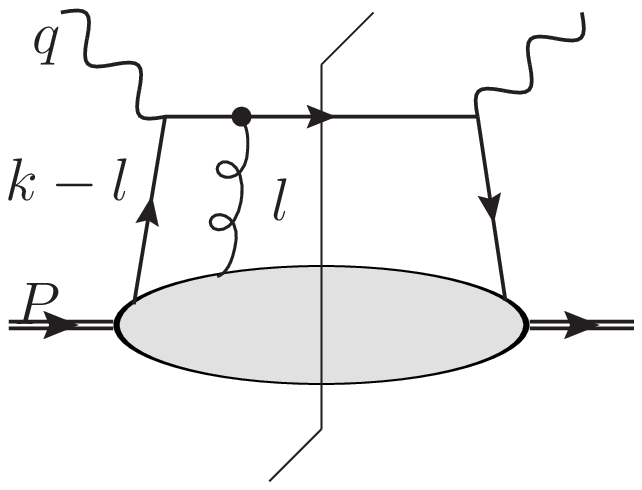}  
    &
      \includegraphics[scale=0.5]{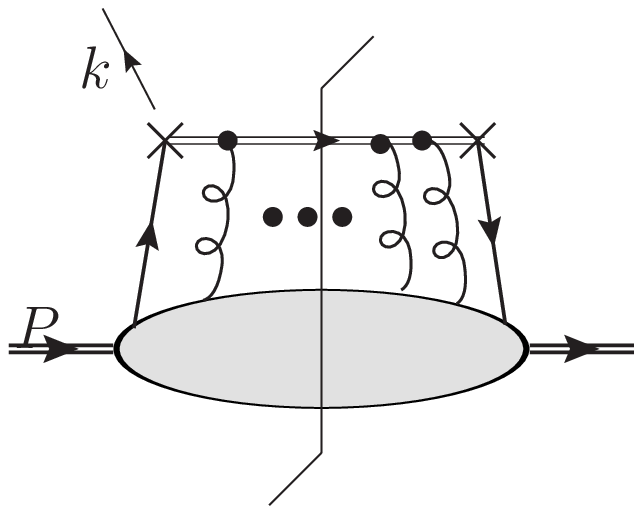}
  \\
     (a) & (b) & (c)
  \end{tabular}

  \caption{(a) Final-state gluon exchanges added to a handbag diagram
    are of leading power.  (b) One-gluon exchange, with momentum
    labeling.  (c) Parton density with Wilson line. }
  \label{fig:handbag.glue}
\end{figure}

We formulate the
gauge condition covariantly as $n\cdot A=0$, where $n^\mu=\delta^\mu_-$.  Then the
gluon propagator has a singularity at $k\cdot n=0$:
\begin{equation}
\label{eq:lcg.prop}
    \frac{i}{k^2} \left( -g^{\mu\nu} + \frac{ k^\mu n^\nu + n^\mu k^\nu  }{ k\cdot n } \right).
\end{equation}
Now, in proving factorization, certain contour deformations are
applied to avoid a rescattering region.  Some relevant graphs are
illustrated in Fig.\ \ref{fig:handbag.glue}(a).  The contour
deformation is away from $k\cdot n\simeq0$, and is obstructed by the gauge
singularity in the light-cone-gauge propagator Eq.\
(\ref{eq:lcg.prop}).

To avoid this problem, we use Feynman gauge.  But the price is that we
have extra regions to consider, notably where the gluons in Fig.\
\ref{fig:handbag.glue}(a) are collinear to the target, a situation
that is power suppressed in light-cone gauge.

\subsection{Wilson lines in integrated parton densities}

The first part of the solution is to make the definition of the
parton densities gauge-invariant, by inserting a Wilson-line operator
between the quark and anti-quark fields.  A Wilson line is a
path-ordered exponential of the gluon field taken along some path
joining the two fields.  The key question is which is the appropriate
path if we are to obtain a parton density that allows the derivation
of a valid factorization property for the cross section.

In the case of an integrated density, (\ref{eq:pdf1.op}), the fields are
separated along exactly the minus direction.  Taking the Wilson line
along the path joining the two fields is appropriate
\cite{Collins:1981uw}, so that we insert into the matrix element a
factor
\begin{equation}
  W(w^-,0)
  =
  P\,  \exp\left({-ig_0 \int_0^{w^-} \diff{y^-} A^+_{(0)\alpha}(0,y^-,\T{0}) t_\alpha } \right),
\end{equation}
which is equal to unity in $A^+=0$ gauge.  Graphs like Fig.\
\ref{fig:handbag.glue}(a) with gluon exchanges between the final-state
quark and everything else, lead to graphs like Fig.\
\ref{fig:handbag.glue}(c) for the parton density, with the double line
indicating the Feynman rules for the Wilson line: The definition
matches the proof.

Supplemented by renormalization of the UV divergences in the parton
density, this definition is satisfactory for integrated parton
densities, as far as is known.  These are the standard parton
densities that are used in much phenomenology.

\subsection{Meaning of Wilson line}

The Wilson line arises from an approximation to the momentum that
flows from a gluon onto a final-state quark.  For example, for
exchange of one gluon of momentum $l$, Fig.\
\ref{fig:handbag.glue}(b), we have
\begin{equation}
\label{eq:coll.approx}
    \frac{1}{(q+k-l)^2-m^2+i\epsilon}
    \simeq \frac{1}{-2(q+k)\cdot l + i\epsilon}
    \simeq \frac{1}{-2q^-l^++i\epsilon}.
\end{equation}
The approximation is valid for a gluon collinear to the target, and
the previously mentioned contour deformation is needed to avoid the
region where $l^+$ gets small with $\T{l}$ fixed.

\begin{figure}
  \centering
   \includegraphics[scale=0.4]{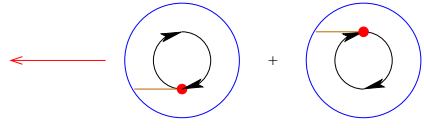}  
   \caption{A quark is knocked out of proton by a virtual photon,
     and acquires acquires a phase accumulated along its path.}
  \label{fig:quark.from.target}
\end{figure}

We interpret this formula by regarding the outgoing quark as having a
diffraction pattern arising from quark emission from different points
in the target, Fig.\ \ref{fig:quark.from.target}.  The quark is highly
time-dilated in the rest frame of the target, and is therefore almost
undeflected while it remains inside the target.  The Wilson line
essentially gives the quantum-mechanical phase acquired by the quark
as it climbs out of gluon field of the target.  The Sivers function, a
transverse-spin-dependent unintegrated parton density that we will
hear about elsewhere in this workshop, occurs because in a spinning
proton, the gluon field rotates and gives different phases on opposite
sides of the proton, just as for optical diffraction from a rotating
transparent ring.

\section{\emph{Unintegrated} pdf and rapidity divergences}

For an unintegrated density (\ref{eq:pdf2}), the parton fields
are no longer at a 
light-like separation, but the derivation of factorization still
requires us to use manipulations like those in Eq.\
(\ref{eq:coll.approx}).  Thus the Wilson line does \emph{not} go along
the line joining the quark and anti-quark; instead, it goes out to
infinity from one field in an approximately light-like direction,
makes a transverse jog, and then comes back to the other field. Thus
the interpretation still applies that the Wilson line gives the phase
acquired by a fast-moving quark going through the gluon field of the
target.  That the Wilson line goes to future infinity, not to past
infinity, is determined by the sign of the $i\epsilon$ in
(\ref{eq:coll.approx}).

With exactly light-like Wilson line, there are divergences at
$l_{\text{gluon}}^+=0$.  For example, for a graph with one exchanged
real or virtual gluon, we have
\begin{subequations}
\label{eq:rap.div}
\begin{align}
\VCeps{scale=0.45,clip}\includegraphics{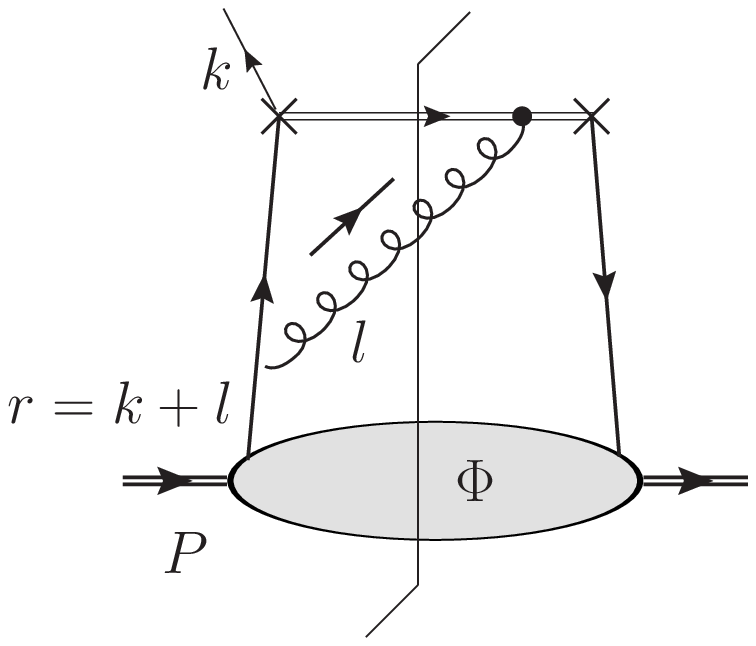}
&~\propto~ 
  \int \diff{r^-} \diff[2]{\T{l}} \int_0 \frac{ \diff{l^+} }{ l^+ }
  ~\frac{ \Phi(k^++l^+,r^-,\T{k}+\T{l}) }{ l_T^2 + m_g^2 ~+~ \cdots  }
\\
\VCeps{scale=0.45,clip}\includegraphics{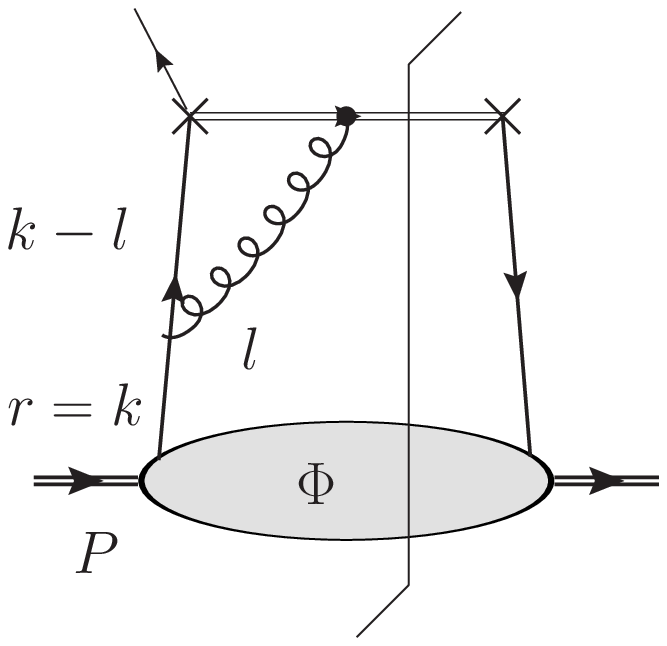}
&~\propto~ 
  ~ - ~ \int \diff{r^-} \diff[2]{\T{l}} \int_0 \frac{ \diff{l^+} }{ l^+ }
  ~ \frac{ \Phi(k^+,r^-,\T{k}) }{ l_T^2 + m_g^2 ~+~ \cdots },
\end{align}
\end{subequations}
where the dots ``$\ldots$'' indicate terms that vanish at $l^+=0$ and that
therefore do not contribute to the divergence.  The two graphs differ
only by placement of the final-state cut.  If we were to integrate
over \emph{all} external transverse momenta $\T{k}$, to get a
contribution to the integrated quark density, the divergence would
cancel.  But in the unintegrated density, the divergence is
uncanceled, and does not cancel against other graphs.  The same
divergence occurs in light-front wave functions --- see, for example,
Eq.\ (29) of \cite{Brodsky:2000ii}.

We included a non-zero gluon mass in Eq.\ (\ref{eq:rap.div}), to
demonstrate clearly that the divergence is not an infra-red
divergence, even though one component of gluon momentum goes to zero.
Equally it cannot be regulated by changing the dimension of
space-time.

\begin{figure}
  \centering
  \standardsubs
  \psfrag{y_n}{\black \small $y_n$}
  \psfrag{y_p}{\blue \small $y_p$}
  \psfrag{collA}{\blue \small coll.\ p}
  \psfrag{pdf}{\blue \small pdf}
  \psfrag{y_J}{\brown \small $y_J$}
  \psfrag{collB}{\brown \small coll.\ J}
  \psfrag{y_Q}{\black \small $y_Q$}
  \psfrag{soft}{\black \small soft}

  \begin{tabular}{c@{\hspace*{1.4cm}}c@{\hspace*{1.4cm}}c}
    \includegraphics[scale=0.6]{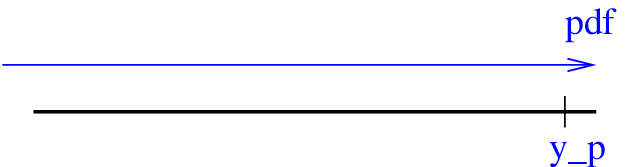}
    &
     \includegraphics[scale=0.6]{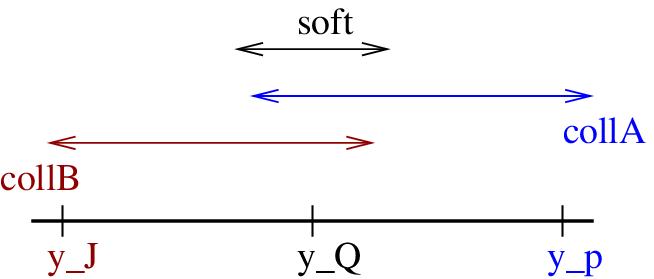}
    &
    \includegraphics[scale=0.6]{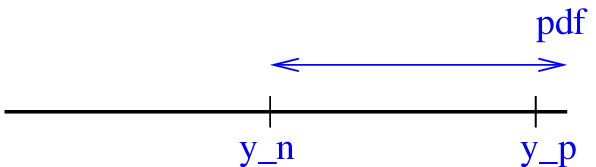}      
  \\
    (a) & (b) & (c)
  \end{tabular}
  
  \caption{Rapidity ranges: 
    (a) for parton density with light-like Wilson line;
    (b) for original integral;
    (c) for parton density with Wilson line of
    rapidity $y_n$.
  }
  \label{fig:rap.range}
\end{figure}

The divergence, in fact, arises from an integral over gluon rapidity,
defined by
\begin{equation}
    y \equiv \frac12 \ln \frac{l^+}{l^-} = \ln \frac{l^+}{ \sqrt{(l^2+l_T^2)/2}}.
\end{equation}
Changing variable from $l^+$ to $y$ at fixed $\T{l}$ gives an
approximately uniform integrand as a function of $y$, between
kinematic limits.  The upper limit is the proton rapidity, but ---
Fig.\ \ref{fig:rap.range}(a) --- the lower limit, given by the Wilson
line, is $-\infty$.

This misrepresents the actual physics, Fig.\ \ref{fig:rap.range}(b).
Only the target-collinear range, approximately between the target and
the virtual photon, belongs in the parton density.  Much lower
rapidities belong with the jet, while a central range should be
associated with a soft factor.  Naturally in a factorization theorem,
it is necessary to compensate double counting of the overlap regions.

\subsection{Modified parton density with cut off on rapidity}

We are therefore required to modify the definition of the parton
density to provide some kind of cut off on gluon rapidity.

One obvious solution is to make the direction $n$ of the Wilson line
non-light-like.  This is essentially the solution of Collins and Soper
\cite{Collins:1981uw,Collins:1981uk}, except that they used a
non-light-like axial gauge $n\cdot A=0$.  As we have already mentioned,
such gauges cause problems in properly deriving factorization, so the
Wilson line solution is preferred.  The definition of a parton density
now has the Wilson line going to $+\infty$ in direction $n$, making a jog
at infinity, and coming back:
\begin{multline}
\label{eq:pdf.def.n}
  f_j(x,\T{k}; y_n) = 
  \int \frac{ \diff{w^-} \diff[2]{\T{w}} }{ (2\pi)^3 } \,
    e^{ -ix P^+w^- + i \T{k}\cdot\T{w} }
\\
    \braket{ p,s| \,\,
            \overline{\psi}(0,w^-,\T{w}) \,
             W(w~{\rm to}~\infty;n)^\dagger \,\,
             \frac{\gamma^+}{2}\,\,
             W(\text{at~} \infty )\,\,
            W(0~{\rm to}~\infty;n) \,
            \psi(0) \,\,
    | p,s }.
\end{multline}
Other variations on the same idea are possible, e.g.,
\cite{Ji:2004wu}, but there must always be present an extra parameter
$y_n$ that limits the gluon rapidity: Fig.\ \ref{fig:rap.range}(c).

Collins, Soper and Sterman (CSS)
\cite{Collins:1981uw,Collins:1981uk,Collins:1984kg} restored the
predictive power of the theory by an equation for the $y_n$
dependence.  The kernel of the equation involves a perturbatively
calculable function $G$, a universal non-perturbative function
$K(q_T)$, and a perturbatively calculable renormalization-group
function $\gamma_K(\alpha_s)$ for $K$ and $G$.

The CSS formalism not only determines a resummation of the large
logarithms of $Q/q_T$ that arise in collinear factorization, but
systematically treats the effects of the non-perturbative region of
small transverse momentum for the partons.


\subsection{Implications}

An immediate consequence of the CSS method is the energy dependence
of the transverse-momentum distribution of the Drell-Yan cross
section.  Fits to the non-perturbative parts of the factorization
formula are made, in principle from a subset of the data, and
predictions then made for all energies, e.g., \cite{Landry:2002ix}.
Extending the formalism to the gluon density allows the transverse
momentum of the Higgs boson to be predicted, e.g.,
\cite{Berger:2002ut}.
Similar methods apply to transverse momentum distributions in other
processes, like semi-inclusive DIS, and almost-back-to-back hadron
production in $e^+e^-$ annihilation.

Close examination of the arguments for the direction of the Wilson
lines shows that in the Drell-Yan process, the Wilson lines must be
point to the past instead of the future \cite{Collins:2004nx}.
Time-reversal symmetry of QCD shows that for most parton densities,
the change in the direction of Wilson line compared with DIS has no
effect on the numerical value of the parton density.  But certain
quantities involving transverse spin change sign \cite{Collins:2002kn}
between DIS and Drell-Yan.  These include the Sivers function, which
governs the transverse-momentum density of partons in a transversely
polarized nucleon, and the Boer-Mulders function, which governs the
transverse polarization of quarks in a unpolarized hadron.

\section{Wilson line self energy?}

\begin{figure}
   \centering
   \psfrag{n}{$n\cdot w$}
   \psfrag{T}{$\T{w}$}
   \psfrag{L}{$L$}
   \VCeps{scale=0.8,clip}\includegraphics{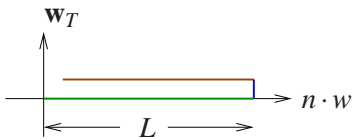}
      
   \caption{Wilson line in unintegrated parton density, viewed from
     side. The limit $L\to\infty$ should be taken. }
   \label{fig:WL.geometry}
\end{figure}

Further complications arise because there is a segment of the Wilson
line that can be regarded as a color dipole extending all the way to
infinity --- Fig.\ \ref{fig:WL.geometry}.  Certain topologies of graph
are required in the parton density, but do not arise in the derivation
of factorization, e.g., in obtaining Fig.\ \ref{fig:handbag.glue}(c)
from Fig.\ \ref{fig:handbag.glue}(a).

\begin{figure}
  \begin{minipage}[t]{0.45\linewidth}
  \centering
   \includegraphics[scale=0.4]{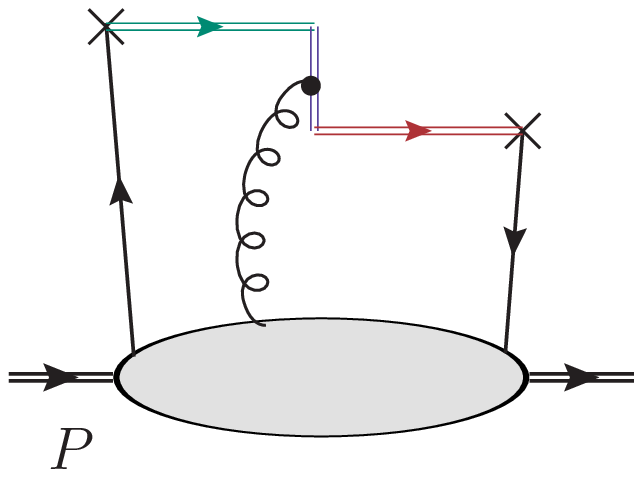}  
   \caption{Gluons joining to the ``jog-at-infinity'' part of the
     Wilson line, color-coded as in Fig.\ \protect\ref{fig:WL.geometry}.}
  \label{fig:to.jog}
  \end{minipage}
\hfill
  \begin{minipage}[t]{0.45\linewidth}
  \centering
   \includegraphics[scale=0.4]{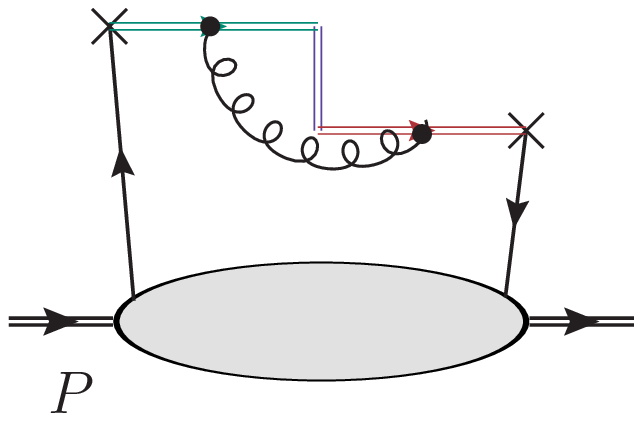}  
   \caption{Self-energy graph for Wilson line.}
  \label{fig:WL.self}
  \end{minipage}
\end{figure}

One complication, found by Belitsky, Ji, and Yuan
\cite{Belitsky:2002sm}, is that definition (\ref{eq:pdf.def.n})
entails graphs like Fig.\ \ref{fig:to.jog} that have gluons connecting
the lower part of the graph to the segment of the Wilson line that
makes the transverse jog at infinity.  These graphs vanish in Feynman
gauge, so they do not affect the proof of factorization.  But they do
appear in the $n\cdot A=0$ gauge, to give gauge-independent parton
densities.  This resolves the problem that otherwise the
Sivers function is zero in axial gauge.

A second complication, not apparently recognized before, gives rise to
extra divergences.  To solve it, I propose here a further redefinition
of the unintegrated quark density to cancel the undesired
contributions.  While this change affects the actual operator
definition, it leaves the CSS evolution equations unaltered.  Thus it
leaves unaltered the phenomenology described in the previous section,
which uses the form of the evolution equation but not the full
implementation of the operator definition.

The divergences were found by Diehl (private communication) using the
Collins-Soper definition in $n\cdot A=0$ gauge.  In the Feynman gauge, the
divergence arises from Wilson line self energies.  The one-gluon case,
Fig.\ \ref{fig:WL.self}, gives a \emph{linear} divergence, which in
coordinate space is proportional to the length $L$ of the Wilson line,
in the limit $L \to \infty$.  This graph does not correspond to any treated
in the factorization proof.

A natural expectation, in need of a full proof, is that to all orders,
and also beyond perturbation theory, the infinitely long dipole will
have $L$-dependence like that of a Wilson loop, i.e., there is an
exponentiation, giving a factor of the form $e^{-LV(w_T)}$ in
coordinate space, where $V(w_T)$ depends on the separation $w_T$
between the two parts of the Wilson line.  This divergence should be
\emph{exactly} the same as in a Wilson loop of size $L\times w_T$.

It is also necessary to cancel graphs where gluons connect the long
sections of the Wilson line to the jog at infinity, including the
associated UV divergences, since none of these graphs appear in the
factorization proof.  Canceling these graphs avoids the problems
discussed in \cite{Cherednikov:2007tw}.

I therefore propose to redefine the unintegrated parton density as
follows:
\begin{equation}
\label{eq:pdf.def.n.final}
  f_j(x,\T{k}; y_n) = 
  \int \frac{ \diff{w^-} \diff[2]{\T{w}} }{ (2\pi)^3 } \,
    e^{ -ix P^+w^- + i \T{k}\cdot\T{w} }
  \quad \lim_{L\to\infty} ~
\frac{ \mbox{Matrix element in (\ref{eq:pdf.def.n})$(L)$} }
{
  \sqrt{\VCeps{scale=0.4,clip}\includegraphics{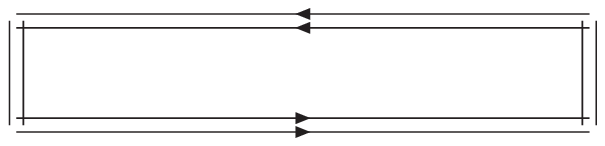}
  (2L)}
} ~ .
\end{equation}
The numerator is just as in Eq.\ (\ref{eq:pdf.def.n}), but with the
Wilson line going out to $n\cdot y = L$ instead of $\infty$.  In the
denominator is a Wilson loop.  Since it has two transverse segments,
whereas the numerator has only one, we take the square root of the
Wilson loop.  Hence, to keep the correct $e^{-LV(w_T)}$ divergence
factor, the Wilson \emph{loop} must be given a length $2L$, when the
numerator has a Wilson line has length $L$.  Finally the limit $L\to\infty$
is taken.

We should now have a satisfactory definition, to be taken literally in
any kind of calculation (perturbative or non-perturbative).  Various
further refinements are possible, e.g., to absorb soft factors in the
CSS factorization formula, but these can be made in terms of the
quantity just defined.

\section{Summary and outlook}

An elementary treatment of DIS in a field theory where the parton
model is valid leads to a definition of a parton density that can be
literally interpreted as an expectation value of a parton number
operator in the sense of light-front quantization.

In QCD the definition must be modified by the insertion of a Wilson
line.  A natural definition is to use a light-like Wilson line.  But
in an unintegrated density, this gives rapidity divergences, which
must be cut-off or avoided somehow, for example by a non-light-like
Wilson line.  The rapidity cut-off $y_n$ is tied to the physics of the
cross section, and there is an equation for the evolution with respect
to $y_n$.

A new result is that further divergences arise from the integral to
spatial infinity of the dipolar Wilson line.  These can be canceled by
a suitable Wilson loop factor, without affecting the standard CSS
phenomenology.  Further work is needed to test whether my conjectures
in this area are correct.

\emph{Note added after presentation of the talk}: The same divergences
infect light-front wave functions.  This presumably necessitates some
changes or distortions in standard formulations of QCD quantized on
the light-front, if the formalism is to be taken literally.


\end{document}